\documentclass{article}

\usepackage[dblblindworkshop, final, nonatbib]{neurips_2025}

\usepackage[utf8]{inputenc} 
\usepackage[T1]{fontenc}    
\usepackage{hyperref}       
\usepackage{url}            
\usepackage{booktabs}       
\usepackage{amsfonts}       
\usepackage{nicefrac}       
\usepackage{microtype}      
\usepackage{xcolor}         
\usepackage{bm}
\usepackage{biblatex}
\usepackage{amsmath}
\usepackage{graphicx}
\usepackage{siunitx}
\usepackage{listings}
\title{Case study of a differentiable heterogeneous multiphysics solver for a nuclear fusion application}
\workshoptitle{DiffSys}

\lstdefinestyle{myStyle}{
    belowcaptionskip=1\baselineskip,
    breaklines=true,
    frame=none,
    numbers=none, 
    basicstyle=\footnotesize\ttfamily,
    keywordstyle=\bfseries\color{blue},
    commentstyle=\itshape\color{purple!40!black},
    backgroundcolor=\color{gray!10!white},
}

\author{%
  Jack B.~Coughlin \\
  Pasteur Labs; Brooklyn, NY \\
  \texttt{jack.coughlin@simulation.science} \\
  \And
  Archis Joglekar \\
  Ergodic LLC; Seattle, WA \\
  Pasteur Labs; Brooklyn, NY \\
  \AND
  Jonathan Brodrick \\
  Pasteur Labs; Brooklyn, NY \\
  \And
  Alexander Lavin \\
  Pasteur Labs \& ISI; Brooklyn, NY \\
}

\begin{document}

\maketitle

\begin{abstract}
This work presents a case study of a heterogeneous multiphysics solver from the nuclear fusion domain. At the macroscopic scale, an auto-differentiable ODE solver in JAX computes the evolution
of the pulsed power circuit and bulk plasma parameters for a compressing Z Pinch. The ODE solver requires a closure for the impedance of the plasma load obtained via root-finding at every timestep, which we solve efficiently using gradient-based Newton iteration. However, incorporating non-differentiable production-grade plasma solvers like Gkeyll (a C/CUDA plasma simulation suite) into a gradient-based workflow is non-trivial. The ``Tesseract'' software addresses this challenge by providing a multi-physics differentiable abstraction layer made fully compatible with JAX (through the \texttt{tesseract\_jax} adapter). This architecture ensures end-to-end differentiability while allowing seamless interchange between high-fidelity solvers (Gkeyll), neural surrogates, and analytical approximations for rapid, progressive prototyping.
\end{abstract}

\section{Introduction}

The deep learning revolution has borne many fruits, but for scientific computing one of the main
benefits has been the development of software frameworks for writing automatically differentiable (AD)
numerical programs.
End-to-end differentiable solvers have been created for fluid dynamics \cite{bezginJAXFluids20HPC2025, franzPICTDifferentiableGPUAccelerated2025},
finite element analysis \cite{xueJAXFEMDifferentiableGPUaccelerated2023}, molecular dynamics \cite{schoenholzJAXMDFramework2020}, and plasma transport in tokamaks \cite{citrinTORAXFastDifferentiable2024}, to name just a few examples.
Differentiable simulations unlock sensitivity analysis and gradient-based optimization workflows 
end-to-end through large PDE solutions.

However, many physical systems of interest in engineering applications are ``multiphysics'' in nature.
By a multiphysics system we mean a system involving a coupling between two or more physical domains or scales,
each governed by its own set of ordinary or partial differential equations. The coupling between equation
is often constrained by an implicit relation that requires the solution of a nonlinear system.
Moreover, spatial and temporal scales may vary widely, necessitating the use of implicit time stepping methods
or convergence to a steady state solution on the fast timescale. Examples of multiphysics systems include
fluid-structure interactions, combusting and reacting flows, and radiation hydrodynamics \cite{Lavin2021SimulationIT}.

Applying AD to multiphysics problems presents a particular challenge
because their complexity may make it infeasible to rewrite all physics solvers in an AD-native framework.
In this work we present one such example. One component of our system is an ODE for which it is quite simple
to write a new solver in JAX. However, this is coupled to a high-dimensional PDE for which an excellent, high-accuracy
and efficient solver is available in C. To leverage this existing battle-tested code we use Tesseracts \cite{hafnerTesseractCoreUniversal2025} to provide an abstraction layer at the boundary between solvers.
Crucially, as Gkeyll itself is non-differentiable, Jacobian-vector products and vector-Jacobian products of the C solver are implemented using finite differences,
and the resulting differentiable function is incorporated into the outer JAX program using \texttt{tesseract\_jax}\footnote{\href{https://github.com/pasteurlabs/tesseract-jax}{https://github.com/pasteurlabs/tesseract-jax}}.
Furthermore, Tesseracts enable a modular software architecture, enabling us to swap out different implementations
for the PDE component, such as neural surrogates or analytic approximations.
End-to-end differentiability of the program is maintained regardless of the underlying solver.

\section{Fusion Z Pinch Compression Modeling}

\subsection{Background}

The Sheared Flow-Stabilized Z Pinch \cite{shumlakShearedFlowStabilizedZPinch2012} is a promising approach to magnetically confined nuclear fusion (MCF).
It has the benefits of a compact size, requires no external magnetic field coils, and relatively simple engineering compared to other MCF concepts.
The Z Pinch functions on the basis of the plasma pinch principle: a plasma carrying a current with density vector $\bm{j}$ produces a magnetic field $\bm{B}$
and corresponding force vector $\bm{F} = \bm{j} \times \bm{B}$ which acts to compress the plasma in on itself.
By ramping up the plasma current, one can increase $\bm{F}$ and thus the equal and opposite pressure gradient force $-\nabla p$, driving the plasma density and temperature
up to fusion conditions.

\subsection{Macroscale circuit equation}

The pulsed power driver of the Z Pinch device can be modeled as a series RLC circuit as done in Ref. \cite{dattaWholeDeviceModeling2024}.
Let $Q$ and $I$ denote the time-dependent capacitor charge and total circuit current respectively. 
The voltage balance law around the circuit, including all three $RLC$ circuit components and the plasma load, is
\begin{align}
    \left( L + L_p + \dot{L_p} \frac{I}{\dot{I}} \right) \dot{I} + R I + \frac{Q}{C} = -V_{R_p},
\end{align}
where $L$ is the circuit inductance, $R$ the circuit resistance and $C$ the capacitance.
The plasma load impedance is characterized by the plasma inductance $L_p$ and the resistive plasma voltage $V_{R_p}$.
The plasma inductance is a geometric property of the pinch profile and is derived in Ref. \cite{crewsKadomtsevPinchRevisited2024a}.
The resistive plasma voltage $V_{R_p}$ is the voltage required to drive a current $I$ through the plasma at steady state,
and is a function of the current $I$ and the plasma temperature $T$ and density $n$.
Temperature $T$ is related to current $I$, linear density $N$, and charge state $Z$ by the well-known Bennett relation for pinches.
The relation between $T$ and $n$ is given by the specific entropy $s = \ln(T / n^{\gamma - 1})$, where $\gamma = 5/3$ is
the adiabatic constant for hydrogenic plasmas.
The complete ODE system for the plasma and circuit state is
\small
\begin{align}
    \label{eqn:Q_I_s}
    \frac{\mathrm{d}}{\mathrm{d}t} \begin{pmatrix}
    Q \\ I \\ s
    \end{pmatrix}
    =
    \begin{pmatrix}
    I \\
    \frac{1}{L + \frac{L_z \mu_0}{2 \pi (\gamma - 1)} + L_p} \left[ - \frac{Q}{C} - R I - V_{R_p}\!(I, T, n) \right] \\
    \frac{\gamma - 1}{(1 + Z) T} (P_\eta - P_{Br})
    \end{pmatrix},
    \quad 
    T = \frac{\mu_0 I^2}{8 \pi (1 + Z) N},
    \quad
    n = \left( \frac{T}{e^s} \right)^{\frac{1}{\gamma-1}}
\end{align}
\normalsize
The terms $P_\eta$ and $P_{Br}$ are the contributions from Ohmic heating of the plasma and bremsstrahlung radiative cooling,
a key energy loss mechanism in fusion plasmas. Expressions for them may be found in \cite{goldstonIntroductionPlasmaPhysics1995a}.

\subsection{Microscale plasma physics closure}

The key remaining challenge is to specify a closure for $V_{R_p}(I, T, n)$.
While the circuit dynamics can be described by the lumped element model eq. \eqref{eqn:Q_I_s}, the plasma impedance depends closely
on the distribution of current-carrying plasma particles near the electrodes.
Plasma-electrode boundaries form a structure known as a plasma sheath, in which the mobility
difference between light electrons and heavy, positively charged ions results in a nonzero charge density at the plasma edge.
It can be shown \cite{stangebyPlasmaBoundaryMagnetic2000} that in the presence of a bias voltage, the plasma current density 
limits to a value that scales as $n \sqrt{T}$. Moreover, computational investigations have shown 
\cite{skolarContinuumKineticInvestigation2023, skolarGeneralKineticIon2025a}
that the current density is sensitive to details of the sheath structure and electrode boundary conditions.
A physically accurate model of the sheath requires the solution of the Vlasov-Poisson-Fokker-Planck (VPFP) system of equations,
\begin{align}
    \label{eqn:vlasov_poisson}
    \partial_t f_s + \bm{v} \cdot \nabla_{\bm{x}} f_s + \frac{q_s}{m_s} \bm{E} \cdot \nabla_{\bm{v}} f_s = \sum_{s'} \nu_{s s'} C_{ss'} + \Gamma_s, \quad \nabla_{\bm{x}} \cdot \bm{E} = \frac{1}{\epsilon_0} \sum_s q_s \int_{\mathbb{R}^3} f_s \, \mathrm{d}\bm{v},
\end{align}
where the species index $s$ varies over the electrons and ions.
In full generality, Eqn. \eqref{eqn:vlasov_poisson} is a 6-dimensional time-dependent partial differential equation (PDE).
Near the magnetic axis of the Z Pinch, we assume spatial symmetry in the azimuthal and radial directions, reducing \eqref{eqn:vlasov_poisson}
to a 2D PDE.
The plasma-electrode boundary condition is a perfectly absorbing wall on both sides of the physical domain.
The term $\Gamma_s$ is a source term that serves to replenish particles and energy lost to the wall.
Formally, the coupling between the VPFP system \eqref{eqn:vlasov_poisson} and the circuit ODE \eqref{eqn:Q_I_s} is
given by
\begin{align}
    \label{eqn:coupling}
    V_{R_p}(I, T, n) = V\; \text{s.t.} \; \begin{cases}
        I = \frac{N}{n} \sum_s q_s \int_{\mathbb{R}^3} \bm{v} f_s^\infty \,\mathrm{d} \bm{v}, \\
        f_s^\infty = f_s(t \rightarrow \infty), \quad \partial_t f_s^\infty = 0, \\
        f_s\vert_{t=0} = \frac{m_s n}{(2\pi T)^{1/2}} \exp \left( -m_s |\bm{v}|^2/(2T) \right), \\
        \bm{E} = \nabla \phi, \quad \phi\vert_{x=L_z} - \phi\vert_{x=0} = V.
    \end{cases}
\end{align}
Equation \eqref{eqn:coupling} is solved via Newton iteration.
A proposed solution $V$ is applied as a Dirichlet boundary condition on the electrostatic
potential $\phi$, and Equation \eqref{eqn:vlasov_poisson} is integrated in time to a steady-state
solution $f_s^\infty$, with initial conditions determined by the plasma temperature and density $T, n$.
The residual is the difference between the target current $I$ and the observed steady-state current.
The Newton iteration requires gradients of the map $V \mapsto I$.

\section{Software architecture}

\begin{figure}
  \centering
  \includegraphics[width=0.85\linewidth]{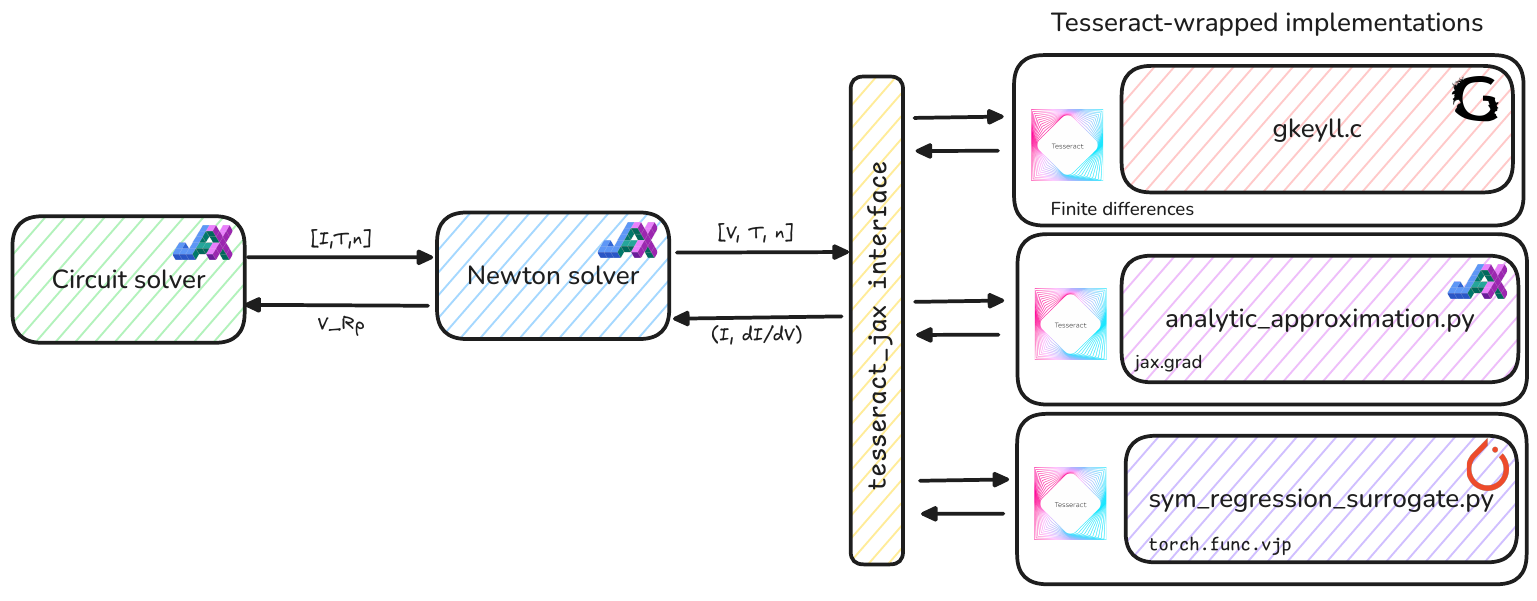}
  \caption{Dataflow diagram of the software components used in solving Equations \eqref{eqn:Q_I_s} and \eqref{eqn:vlasov_poisson}.
      Different implementations of the $V \mapsto I$ map are packaged in Tesseracts. Implementation details are hidden behind
  a \texttt{tesseract\_jax} call enabling rapid experimentation with different models. \label{fig:arch}}
\end{figure}

Our implementation uses JAX \cite{jax2018github}, a Python framework for writing AD programs.
Equation \eqref{eqn:Q_I_s} is discretized using Diffrax \cite{kidger2021on}, and the Newton inversion of the $V \mapsto I$ map
is implemented with the Optimistix library \cite{optimistix2024}.

%
We have three implementations of the V→I map occupying 
different points on the accuracy-efficiency curve: 
(1) The C/CUDA suite Gkeyll provides high-fidelity solutions requiring ~18 GPU-hours per trajectory, but
gradients can only be approximated using finite differences; (2) a symbolic
regression trained on the outputs of a purpose-built semi-Lagrangian VPFP solver, with AD provided by PyTorch; (3) a pure JAX closed-form approximation to the saturation current  
\cite{skolarContinuumKineticInvestigation2023} to enable rapid prototyping
The architecture is diagrammed in Figure \ref{fig:arch}.
The Tesseract abstraction enables seamless interchange between these models without modifying the outer optimization loop (Listing \ref{newton}) and ODE solve.

\section{Results}

\begin{figure}
  \centering
  \includegraphics[width=0.5\linewidth]{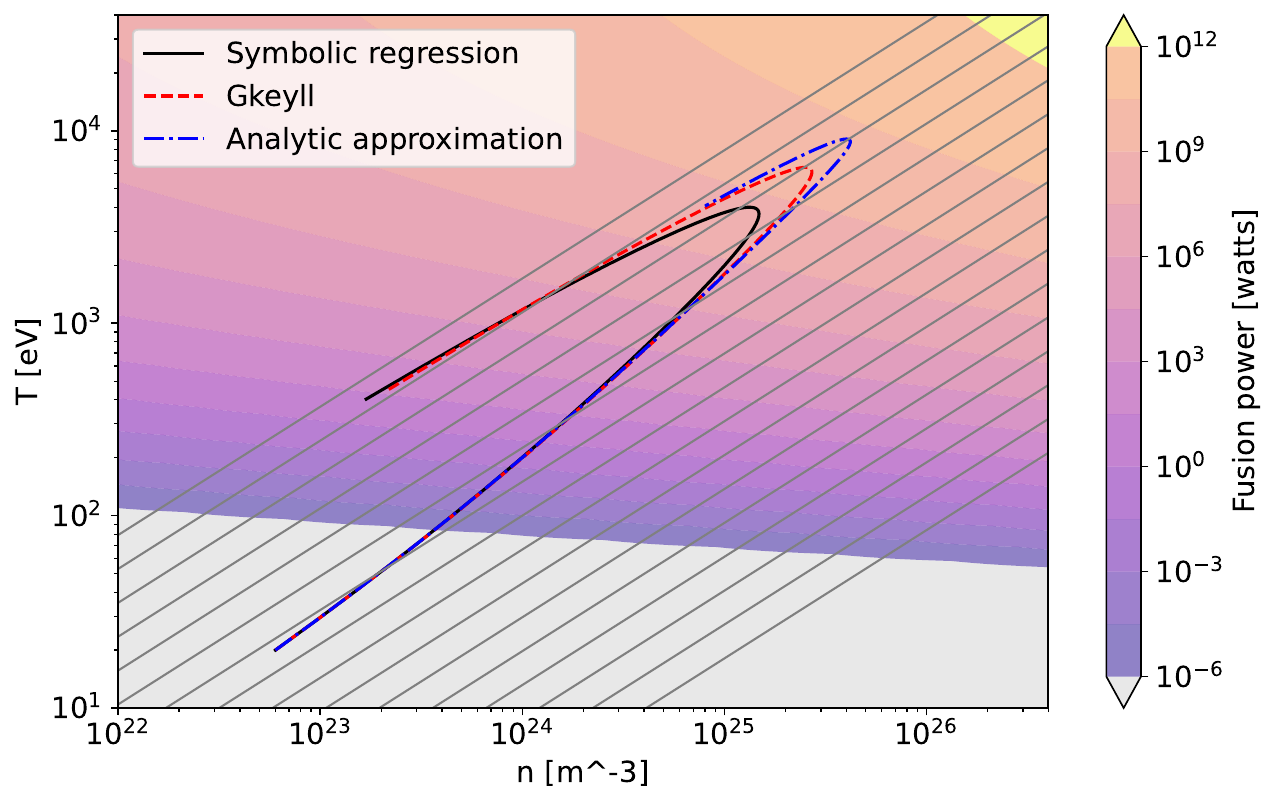}
  \caption{Density-temperature trajectories of three rollouts of the circuit ODE, each with a different plasma model.
  Straight lines are lines of constant specific entropy.
    \label{fig:forward_runs}}
  \includegraphics[width=0.8\linewidth]{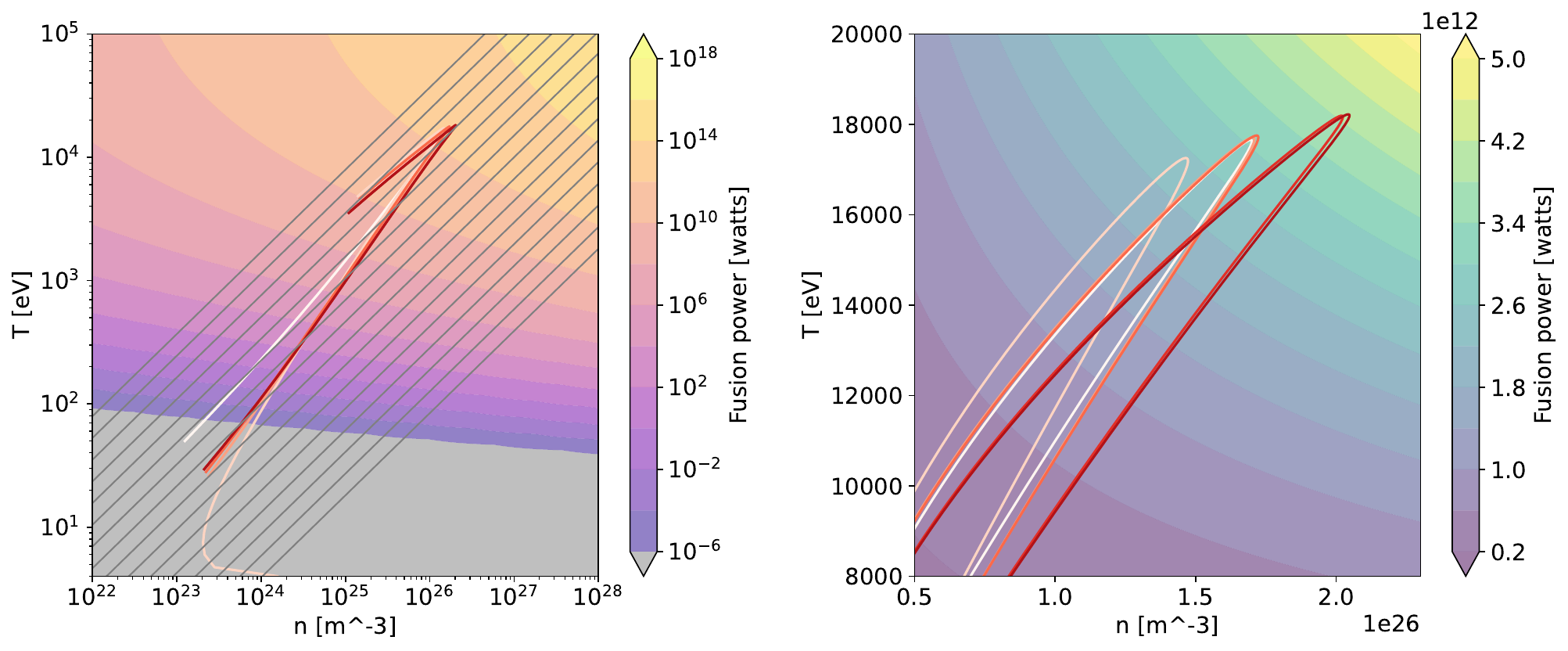}
  \caption{Density-temperature trajectories of successive optimization iterates. We optimize over initial temperature $T$ and circuit capacitance $C$ using the L-BFGS quasi-Newton method. The gradient-based optimizer quickly traverses the two-dimensional parameter space.
    \label{fig:optimization}}
\begin{lstlisting}[style=myStyle, language=Python, caption=Pseudocode for JAX implementation of a Newton solve through an arbitrary Tesseract., label=newton]
def newton_solve(I, T, n, plasma_tesseract):
    def residual(V_guess, args):
        return tesseract_jax.apply_tesseract(
            plasma_tesseract, {"V": V_guess, "T": T, "n": n}
        ) - I
    solver = optimistix.Newton(rtol=1e-4, atol=1e-4)
    return optimistix.root_find(residual, T, solver)
\end{lstlisting}
\end{figure}

To demonstrate the flexibility of the implementation, Figure \ref{fig:forward_runs} presents solution trajectories in density-temperature space for the three different plasma models.
The solver is end-to-end differentiable for any choice of plasma model Tesseract.
Figure \ref{fig:optimization} presents the solution trajectories of successive optimization iterates for the closed form approximation. We use L-BFGS to optimize for $Q$, the ratio of fusion energy to capacitor energy, with respect to initial temperature $T\vert_{t=0}$ and capacitance $C$.

\section{Conclusion}
Multiphysics simulations in engineering applications often require 
coupling heterogeneous software components including legacy C/CUDA 
codes that cannot be easily rewritten in AD frameworks. We 
demonstrated that Tesseracts enable end-to-end differentiable 
multiphysics workflows that incorporate such solvers without dealing with low-level
custom JAX primitives or code rewrites. The key benefit is 
modularity: our implementation seamlessly interchanges between 
high-fidelity (Gkeyll), surrogate (symbolic regression), and 
analytical closure models, enabling rapid design-space exploration 
with appropriate accuracy-efficiency tradeoffs. This approach is 
particularly valuable when computational budgets vary (e.g., 
hyperparameter sweeps vs. final validation) or when solver 
fidelity requirements are uncertain during development.

\section*{Acknowledgements and Disclosure of Funding}

This research used resources of the National Energy Research Scientific Computing Center, a DOE Office of Science User Facility supported by the Office of Science of the U.S. Department of Energy under Contract No. DE-AC02-05CH11231 using NERSC award FES-ERCAP0026741. This work also received support from Pasteur Labs and Institute for Simulation Intelligence (ISI).  The authors declare no competing interests.

\printbibliography

@article{bezginJAXFluids20HPC2025,
  title = {{{JAX-Fluids}} 2.0: {{Towards HPC}} for Differentiable {{CFD}} of Compressible Two-Phase Flows},
  shorttitle = {{{JAX-Fluids}} 2.0},
  author = {Bezgin, Deniz A. and Buhendwa, Aaron B. and Adams, Nikolaus A.},
  year = 2025,
  month = mar,
  journal = {Computer Physics Communications},
  volume = {308},
  pages = {109433},
  issn = {00104655},
  doi = {10.1016/j.cpc.2024.109433},
  urldate = {2025-02-20},
  langid = {english}
}

@misc{citrinTORAXFastDifferentiable2024,
  title = {{{TORAX}}: {{A Fast}} and {{Differentiable Tokamak Transport Simulator}} in {{JAX}}},
  shorttitle = {{{TORAX}}},
  author = {Citrin, Jonathan and Goodfellow, Ian and Raju, Akhil and Chen, Jeremy and Degrave, Jonas and Donner, Craig and Felici, Federico and Hamel, Philippe and Huber, Andrea and Nikulin, Dmitry and Pfau, David and Tracey, Brendan and Riedmiller, Martin and Kohli, Pushmeet},
  year = 2024,
  month = dec,
  number = {arXiv:2406.06718},
  eprint = {2406.06718},
  primaryclass = {physics},
  publisher = {arXiv},
  doi = {10.48550/arXiv.2406.06718},
  urldate = {2025-10-09},
  archiveprefix = {arXiv}
}

@article{crewsKadomtsevPinchRevisited2024a,
  title = {The {{Kadomtsev}} Pinch Revisited for Sheared-Flow-Stabilized {{Z-pinch}} Modeling},
  author = {Crews, Daniel W. and Datta, Iman A. M. and Meier, Eric T. and Shumlak, Uri},
  year = 2024,
  month = oct,
  journal = {IEEE Transactions on Plasma Science},
  volume = {52},
  number = {10},
  eprint = {2404.06636},
  primaryclass = {physics},
  pages = {4804--4816},
  issn = {0093-3813, 1939-9375},
  doi = {10.1109/TPS.2024.3383312},
  urldate = {2025-05-23},
  archiveprefix = {arXiv},
  langid = {english}
}

@article{dattaWholeDeviceModeling2024,
  title = {Whole Device Modeling of the Fuze Sheared-Flow-Stabilized {{Z}} Pinch},
  author = {Datta, I.A.M. and Meier, E.T. and Shumlak, U.},
  year = 2024,
  month = jun,
  journal = {Nuclear Fusion},
  volume = {64},
  number = {6},
  pages = {066016},
  issn = {0029-5515, 1741-4326},
  doi = {10.1088/1741-4326/ad3fcb},
  urldate = {2025-06-03},
  langid = {english}
}

@misc{franzPICTDifferentiableGPUAccelerated2025,
  title = {{{PICT}} -- {{A Differentiable}}, {{GPU-Accelerated Multi-Block PISO Solver}} for {{Simulation-Coupled Learning Tasks}} in {{Fluid Dynamics}}},
  author = {Franz, Aleksandra and Wei, Hao and Guastoni, Luca and Thuerey, Nils},
  year = 2025,
  month = may,
  number = {arXiv:2505.16992},
  eprint = {2505.16992},
  primaryclass = {cs},
  publisher = {arXiv},
  doi = {10.48550/arXiv.2505.16992},
  urldate = {2025-10-09},
  archiveprefix = {arXiv}
}

@book{goldstonIntroductionPlasmaPhysics1995a,
  title = {Introduction to Plasma Physics},
  author = {Goldston, R. J. and Rutherford, P. H.},
  year = 1995,
  publisher = {Institute of Physics Pub},
  address = {Bristol, UK ; Philadelphia},
  isbn = {978-0-7503-0325-5 978-0-7503-0183-1},
  langid = {english},
  lccn = {QC718 .G63 1995}
}

@article{hafnerTesseractCoreUniversal2025,
  title = {Tesseract {{Core}}: {{Universal}}, Autodiff-Native Software Components for {{Simulation Intelligence}}},
  shorttitle = {Tesseract {{Core}}},
  author = {H{\"a}fner, Dion and Lavin, Alexander},
  year = 2025,
  month = jul,
  journal = {Journal of Open Source Software},
  volume = {10},
  number = {111},
  pages = {8385},
  issn = {2475-9066},
  doi = {10.21105/joss.08385},
  urldate = {2025-10-09},
  copyright = {http://creativecommons.org/licenses/by/4.0/}
}

@misc{jax2018github,
  title = {{{JAX}}: Composable Transformations of {{Python}}+{{NumPy}} Programs},
  author = {Bradbury, James and Frostig, Roy and Hawkins, Peter and Johnson, Matthew James and Leary, Chris and Maclaurin, Dougal and Necula, George and Paszke, Adam and VanderPlas, Jake and {Wanderman-Milne}, Skye and Zhang, Qiao},
  year = 2018
}

@phdthesis{kidger2021on,
  title = {On {{Neural Differential Equations}}},
  author = {Kidger, Patrick},
  year = 2021,
  school = {University of Oxford}
}

@article{Lavin2021SimulationIT,
  title={Simulation Intelligence: Towards a New Generation of Scientific Methods},
  author={Alexander Lavin and Hector Zenil and Brooks Paige and David C. Krakauer and Justin Emile Gottschlich and Timothy G. Mattson and Anima Anandkumar and Sanjay Choudry and Kamil Rocki and Atilim Gunecs Baydin and Carina Prunkl and Olexandr Isayev and Erik J Peterson and Peter Leonard McMahon and Jakob H. Macke and Kyle Cranmer and Jiaxin Zhang and Haruko Murakami Wainwright and Adi Hanuka and Manuela M. Veloso and Samuel A. Assefa and Stephan Zheng and Avi Pfeffer},
  journal={ArXiv},
  year={2021},
  volume={abs/2112.03235},
  url={https://api.semanticscholar.org/CorpusID:244909059}
}

@article{optimistix2024,
  title = {Optimistix: Modular Optimisation in {{JAX}} and Equinox},
  author = {Rader, Jason and Lyons, Terry and Kidger, Patrick},
  year = 2024,
  journal = {arXiv:2402.09983},
  eprint = {2402.09983},
  archiveprefix = {arXiv}
}

@misc{schoenholzJAXMDFramework2020,
  title = {{{JAX}}, {{M}}.{{D}}.: {{A Framework}} for {{Differentiable Physics}}},
  shorttitle = {{{JAX}}, {{M}}.{{D}}.},
  author = {Schoenholz, Samuel S. and Cubuk, Ekin D.},
  year = 2020,
  month = dec,
  number = {arXiv:1912.04232},
  eprint = {1912.04232},
  primaryclass = {physics},
  publisher = {arXiv},
  doi = {10.48550/arXiv.1912.04232},
  urldate = {2025-10-09},
  archiveprefix = {arXiv}
}

@article{shumlakShearedFlowStabilizedZPinch2012,
  title = {The {{Sheared-Flow Stabilized Z-Pinch}}},
  author = {Shumlak, U. and Chadney, J. and Golingo, R.P. and Den Hartog, D.J. and Hughes, M.C. and Knecht, S.D. and Lowrie, W. and Lukin, V.S. and Nelson, B.A. and Oberto, R.J. and Rohrbach, J.L. and Ross, M.P. and Vogman, G.V.},
  year = 2012,
  month = jan,
  journal = {Fusion Science and Technology},
  volume = {61},
  number = {1T},
  pages = {119--124},
  issn = {1536-1055, 1943-7641},
  doi = {10.13182/FST12-A13407},
  urldate = {2024-09-24},
  langid = {english}
}

@article{skolarContinuumKineticInvestigation2023,
  title = {Continuum Kinetic Investigation of the Impact of Bias Potentials in the Current Saturation Regime on Sheath Formation},
  author = {Skolar, Chirag R. and Bradshaw, Kolter and Juno, James and Srinivasan, Bhuvana},
  year = 2023,
  month = jan,
  journal = {Physics of Plasmas},
  volume = {30},
  number = {1},
  eprint = {2211.06488},
  primaryclass = {physics},
  pages = {012504},
  issn = {1070-664X, 1089-7674},
  doi = {10.1063/5.0134656},
  urldate = {2025-05-23},
  archiveprefix = {arXiv},
  langid = {english}
}

@misc{skolarGeneralKineticIon2025a,
  title = {General Kinetic Ion Induced Electron Emission Model for Metallic Walls Applied to Biased {{Z-pinch}} Electrodes},
  author = {Skolar, Chirag R. and Bradshaw, Kolter and Francisquez, Manaure and Murillo, Lucio and Kumar, Vignesh Krishna and Srinivasan, Bhuvana},
  year = 2025,
  month = feb,
  number = {arXiv:2502.01802},
  eprint = {2502.01802},
  primaryclass = {physics},
  publisher = {arXiv},
  doi = {10.48550/arXiv.2502.01802},
  urldate = {2025-05-23},
  archiveprefix = {arXiv},
  langid = {english}
}

@book{stangebyPlasmaBoundaryMagnetic2000,
  title = {The Plasma Boundary of Magnetic Fusion Devices},
  author = {Stangeby, P. C.},
  year = 2000,
  series = {Plasma Physics Series},
  publisher = {Institute of Physics Pub},
  address = {Bristol ; Philadelphia},
  isbn = {978-0-7503-0559-4},
  langid = {english},
  lccn = {QC718.5.C65 S725 2000}
}

@article{xueJAXFEMDifferentiableGPUaccelerated2023,
  title = {{{JAX-FEM}}: {{A}} Differentiable {{GPU-accelerated 3D}} Finite Element Solver for Automatic Inverse Design and Mechanistic Data Science},
  shorttitle = {{{JAX-FEM}}},
  author = {Xue, Tianju and Liao, Shuheng and Gan, Zhengtao and Park, Chanwook and Xie, Xiaoyu and Liu, Wing Kam and Cao, Jian},
  year = 2023,
  month = oct,
  journal = {Computer Physics Communications},
  volume = {291},
  pages = {108802},
  issn = {00104655},
  doi = {10.1016/j.cpc.2023.108802},
  urldate = {2025-10-09},
  langid = {english}
}

\appendix
\newpage
\section{Technical Appendices}

\subsection{Simulation parameters}

Details of the numerical setup of the Vlasov-Poisson system match those used in \cite{skolarContinuumKineticInvestigation2023}, except for $T$ and $n$ which take their values from the ODE solver.

\begin{table}[h]
    \caption{Simulation parameters for rollouts contained in Figure \ref{fig:forward_runs}}
  \centering
  \begin{tabular}{lll}
    \toprule
    Parameter     & Description     & Size ($\mu$m) \\
    \midrule
    $C$ & Capacitance & \qty{400}{\micro\farad} \\
    $R$     & Circuit resistance & \qty{1.5}{m\ohm} \\
    $L$     & Circuit inductance       & \qty{200}{nH}  \\
    $V_c$   & Initial capacitor voltage & \qty{40}{kV} \\
    $T\vert_{t=0}$  & Initial temperature & \qty{20}{eV} \\
    $n\vert_{t=0}$  & Initial density & \qty{6e22}{m^{-3}} \\
    $L_z$  & Axial length & \qty{50}{cm} \\
    $r_w$  & Outer electrode radius & \qty{10}{cm} \\
    $Z$  & Charge state & 1 \\
    $N$  & Linear density & \qty{1e19}{m^{-1}} \\
    \bottomrule
  \end{tabular}
\end{table}

\begin{table}[h]
  \caption{Simulation parameters for optimization problem contained in Figure \ref{fig:optimization}}
  \centering
  \begin{tabular}{lll}
    \toprule
    Parameter     & Description     & Size ($\mu$m) \\
    \midrule
    $C$ & Capacitance & $\qty{100}{\micro\farad} \leq C \leq \qty{400}{\micro\farad}$ \\
    $R$     & Circuit resistance & \qty{1.5}{m\ohm} \\
    $L$     & Circuit inductance       & \qty{286}{nH}  \\
    $V_c$   & Initial capacitor voltage & $\qty{0}{V} \leq V_c \leq \qty{80}{kV}$ \\
    $T\vert_{t=0}$  & Initial temperature & $\qty{10}{eV} \leq T \leq \qty{1}{keV}$ \\
    $n\vert_{t=0}$  & Initial density & $n = \qty{1}{\mega\pascal} / T$ \\
    $L_z$  & Axial length & \qty{50}{cm} \\
    $r_w$  & Outer electrode radius & \qty{10}{cm} \\
    $Z$  & Charge state & 1 \\
    $N$  & Linear density & \qty{1e19}{m^{-1}} \\
    \bottomrule
  \end{tabular}
\end{table}

\subsection{Computing resources}

The Gkeyll run included in Figure \ref{fig:forward_runs} required 18 hours on a single A100 GPU on the Perlmutter cluster. The other runs were done on a 2024 M4 Macbook Pro.

\end{document}